\documentclass[sigconf]{acmart}

\hypersetup{draft}

\usepackage{flushend} 

\usepackage{bm}
\usepackage{booktabs}
\usepackage[font=scriptsize]{subcaption}
\usepackage[font=footnotesize]{caption}

\usepackage[nowarn,acronyms,nonumberlist,nopostdot,nomain,nogroupskip]{glossaries}
\glsdisablehyper

\newacronym{3gpp}{3GPP}{3rd Generation Partnership Project}
\newacronym{adc}{ADC}{Analog to Digital Converter}
\newacronym{5g}{5G}{5th generation}
\newacronym{6g}{6G}{6th generation}
\newacronym{aimd}{AIMD}{Additive Increase Multiplicative Decrease}
\newacronym{am}{AM}{Acknowledged Mode}
\newacronym{amc}{AMC}{Adaptive Modulation and Coding}
\newacronym{aqm}{AQM}{Active Queue Management}
\newacronym{awgn}{AGWN}{Additive White Gaussian Noise}
\newacronym{balia}{BALIA}{Balanced Link Adaptation}
\newacronym{bdp}{BDP}{Bandwidth-Delay Product}
\newacronym{qos}{QoS}{Quality of Service}
\newacronym{qoe}{QoE}{Quality of Experience}
\newacronym{pqos}{PQoS}{Predictive Quality of Service}
\newacronym{bf}{BF}{Beamforming}
\newacronym{cc}{CC}{Congestion Control}
\newacronym{cu}{CU}{Centralized Unit}
\newacronym{ai}{AI}{artificial intelligence}
\newacronym{dql}{DQL}{Double Q-learning}
\newacronym{du}{DU}{Distributed Unit}
\newacronym{cdf}{CDF}{Cumulative Distribution Function}
\newacronym{lidar}{LiDAR}{Light Detection and Ranging}
\newacronym{cn}{CN}{Core Network}
\newacronym{rl}{RL}{Reinforcement Learning}
\newacronym{cam}{CAM}{Cooperative Awareness Message}
\newacronym{mse}{MSE}{Mean Squared Error}
\newacronym{cqi}{CQI}{Channel Quality Information}
\newacronym[firstplural=Markov Decision Processes (MDPs)]{mdp}{MDP}{Markov Decision Process}
\newacronym{cp}{CP}{Control Plane}
\newacronym{csirs}{CSI-RS}{Channel State Information - Reference Signal}
\newacronym{dc}{DC}{Dual Connectivity}
\newacronym{dce}{DCE}{Direct Code Execution}
\newacronym{dci}{DCI}{Downlink Control Information}
\newacronym{dl}{DL}{Downlink}
\newacronym{dmr}{DMR}{Deadline Miss Ratio}
\newacronym{dmrs}{DMRS}{DeModulation Reference Signal}
\newacronym{e2e}{E2E}{end-to-end}
\newacronym{ecn}{ECN}{Explicit Congestion Notification}
\newacronym{edf}{EDF}{Earliest Deadline First}
\newacronym{enb}{eNB}{evolved Node Base}
\newacronym{epc}{EPC}{Evolved Packet Core}
\newacronym{es}{ES}{Edge Server}
\newacronym{fdma}{FDMA}{Frequency Division Multiple Access}
\newacronym{fdd}{FDD}{Frequency Division Duplexing}
\newacronym[firstplural=Radio Access Technologies (RATs)]{rat}{RAT}{Radio Access Technology}
\newacronym{fs}{FS}{Fast Switching}
\newacronym{ftp}{FTP}{File Transfer Protocol}
\newacronym{gnb}{gNB}{Next Generation Node Base}
\newacronym{harq}{HARQ}{Hybrid Automatic Repeat reQuest}
\newacronym{hetnet}{HetNet}{Heterogeneous Network}
\newacronym{hh}{HH}{Hard Handover}
\newacronym{hol}{HOL}{Head-of-Line}
\newacronym{ia}{IA}{Initial Access}
\newacronym{ieee}{IEEE}{Institute of Electrical and Electronics Engineers}
\newacronym{imt}{IMT}{International Mobile Telecommunication}
\newacronym{iot}{IoT}{Internet of Things}
\newacronym{ldpc}{LDPC}{Low-Density Parity Check}
\newacronym{los}{LOS}{Line-of-Sight}
\newacronym{lte}{LTE}{Long Term Evolution}
\newacronym{m2m}{M2M}{Machine to Machine}
\newacronym{ml}{ML}{machine learning}
\newacronym{mac}{MAC}{Medium Access Control}
\newacronym{mc}{MC}{Multi-Connectivity}
\newacronym{mcs}{MCS}{Modulation and Coding Scheme}
\newacronym{mec}{MEC}{Mobile Edge Cloud}
\newacronym{mi}{MI}{Mutual Information}
\newacronym{mimo}{MIMO}{Multiple Input, Multiple Output}
\newacronym{mmwave}{mmWave}{millimeter wave}
\newacronym{mptcp}{MPTCP}{Multipath TCP}
\newacronym{mr}{MR}{Maximum Rate}
\newacronym{mss}{MSS}{Maximum Segment Size}
\newacronym{mtd}{MTD}{Machine-Type Device}
\newacronym{mtu}{MTU}{Maximum Transmission Unit}
\newacronym{nfv}{NFV}{Network Function Virtualization}
\newacronym{nlos}{NLOS}{Non-Line-of-Sight}
\newacronym{nlosv}{NLOSv}{Vehicle Non-Line-of-Sight}
\newacronym{nr}{NR}{New Radio}
\newacronym{ofdm}{OFDM}{Orthogonal Frequency Division Multiplexing}
\newacronym{pdcch}{PDCCH}{Physical Downlonk Control Channel}
\newacronym{pdcp}{PDCP}{Packet Data Convergence Protocol}
\newacronym{pdsch}{PDSCH}{Physical Downlink Shared Channel}
\newacronym{pdu}{PDU}{Packet Data Unit}
\newacronym{pf}{PF}{Proportional Fair}
\newacronym{pgw}{PGW}{Packet Gateway}
\newacronym{phy}{PHY}{Physical}
\newacronym{pbch}{PBCH}{Physical Broadcast Channel}
\newacronym[plural=\gls{mme}s,firstplural=Mobility Management Entities (MMEs)]{mme}{MME}{Mobility Management Entity}
\newacronym{prb}{PRB}{Physical Resource Block}
\newacronym{pss}{PSS}{Primary Synchronization Signal}
\newacronym{pscch}{PSCCH}{Physical Sidelink Control Channel}
\newacronym{pucch}{PUCCH}{Physical Uplink Control Channel}
\newacronym{pusch}{PUSCH}{Physical Uplink Shared Channel}
\newacronym{rach}{RACH}{Random Access Channel}
\newacronym{ran}{RAN}{Radio Access Network}
\newacronym{red}{RED}{Random Early Detection}
\newacronym{rf}{RF}{Radio Frequency}
\newacronym{rlc}{RLC}{Radio Link Control}
\newacronym{rlf}{RLF}{Radio Link Failure}
\newacronym{rrc}{RRC}{Radio Resource Control}
\newacronym{rrm}{RRM}{Radio Resource Management}
\newacronym{rru}{RRU}{Remote Radio Unit}
\newacronym{rr}{RR}{Round Robin}
\newacronym{rs}{RS}{Remote Server}
\newacronym{rsrp}{RSRP}{Reference Signal Received Power}
\newacronym{rss}{RSS}{Received Signal Strength}
\newacronym{rtt}{RTT}{Round Trip Time}
\newacronym{rw}{RW}{Receive Window}
\newacronym{rx}{RX}{Receiver}
\newacronym{sa}{SA}{standalone}
\newacronym{sack}{SACK}{Selective Acknowledgment}
\newacronym{sap}{SAP}{Service Access Point}
\newacronym{sc}{SC}{Single Carrier}
\newacronym{sch}{SCH}{Secondary Cell Handover}
\newacronym{scoot}{SCOOT}{Split Cycle Offset Optimization Technique}
\newacronym{sdma}{SDMA}{Spatial Division Multiple Access}
\newacronym{sinr}{SINR}{Signal to Interference plus Noise Ratio}
\newacronym{sl}{SL}{Sidelink}
\newacronym{sm}{SM}{Saturation Mode}
\newacronym{snr}{SNR}{Signal-to-Noise-Ratio}
\newacronym{son}{SON}{Self-Organizing Network}
\newacronym{ss}{SS}{Synchronization Signal}
\newacronym{srs}{SRS}{Sounding Reference Signal}
\newacronym{sss}{SSS}{Secondary Synchronization Signal}
\newacronym{tb}{TB}{Transport Block}
\newacronym{tcp}{TCP}{Transmission Control Protocol}
\newacronym{tdd}{TDD}{Time Division Duplexing}
\newacronym{tdma}{TDMA}{Time Division Multiple Access}
\newacronym{tfl}{TfL}{Transport for London}
\newacronym{tm}{TM}{Transparent Mode}
\newacronym{trp}{TRP}{Transmitter Receiver Pair}
\newacronym{tti}{TTI}{Transmission Time Interval}
\newacronym{ttt}{TTT}{Time-to-Trigger}
\newacronym{tx}{TX}{Transmitter}
\newacronym{ue}{UE}{User Equipment}
\newacronym{ul}{UL}{Uplink}
\newacronym{uml}{UML}{Unified Modeling Language}
\newacronym{um}{UM}{Unacknowledged Mode}
\newacronym{utc}{UTC}{Urban Traffic Control}
\newacronym{vm}{VM}{Virtual Machine}
\newacronym{rsrq}{RSRQ}{Reference Signal Received Quality}
\newacronym{rssi}{RSSI}{Received Signal Strength Indicator}
\newacronym{crs}{CRS}{Cell Reference Signal}
\newacronym{nsa}{NSA}{Non Stand Alone}
\newacronym{mrdc}{MR-DC}{Multi \gls{rat} \gls{dc}}
\newacronym{endc}{EN-DC}{E-UTRAN-\gls{nr} \gls{dc}}
\newacronym{5gc}{5GC}{5G Core}
\newacronym{si}{SI}{Study Item}
\newacronym{iab}{IAB}{Integrated Access and Backhaul}
\newacronym{wf}{WF}{Wired-first}
\newacronym{hqf}{HQF}{Highest-quality-first}
\newacronym{pa}{PA}{Position-aware}
\newacronym{mlr}{MLR}{Maximum-local-rate}
\newacronym{wbf}{WBF}{Wired Bias Function}
\newacronym{mib}{MIB}{Master Information Block}
\newacronym{sib}{SIB}{Secondary Information Block}
\newacronym{rnti}{RNTI}{Radio Network Temporary Identifier}
\newacronym{dft}{DFT}{Discrete Fourier Transform}
\newacronym{kpi}{KPI}{Key Performance Indicator}
\newacronym{ppp}{PPP}{Poisson Point Process}
\newacronym{v2v}{V2V}{Vehicle-to-Vehicle}
\newacronym{wave}{WAVE}{Wireless Access in Vehicular Environments}
\newacronym{udp}{UDP}{User Datagram Protocol}
\newacronym{upa}{UPA}{Uniform Planar Array}
\newacronym{fec}{FEC}{Forward Error Correction}
\newacronym{v2x}{V2X}{Vehicle-To-Everything}
\newacronym{psfch}{PSFCH}{Physical Sidelink Feedback Channel}
\newacronym{pssch}{PSSCH}{Physical Sidelink Shared Channel}
\newacronym{csma}{CSMA}{Carrier Sense Multiple Access}
\newacronym{v2n}{V2N}{Vehicle-to-Network}
\newacronym{wlan}{WLAN}{Wireless Local Area Network}
\newacronym{cav}{CAV}{Connected and Autonomous Vehicle}
\newacronym{v2i}{V2I}{Vehicle-to-Infrastructure}
\newacronym{d2d}{D2D}{Device-to-Device}
\newacronym{c-its}{C-ITS}{Connected Intelligent Transportation System}
\newacronym{fr2}{FR2}{Frequency Range 2}
\newacronym{fr1}{FR1}{Frequency Range 1}
\newacronym{bs}{BS}{Base Station}
\newacronym{sdu}{SDU}{Service Data Unit}
\newacronym{csi}{CSI}{Channel State Information}
\newacronym{scs}{SCS}{Subcarrier Spacing}
\newacronym{sumo}{SUMO}{Simulation of Urban MObility}
\newacronym{prr}{PRR}{Packet Reception Ratio}
\newacronym{dnn}{DNN}{Deep Neural Network}
\newacronym{pdr}{PDR}{Packet Delivery Ratio}
\newacronym{edca}{EDCA}{Enhanced Distribution Channel Access}
\newacronym{sdap}{SDAP}{Service Data Adaptation Protocol}
\newacronym{osm}{OSM}{OpenStreetMap}
\newacronym{rsu}{RSU}{Road Side Unit}
\newacronym{hv}{HV}{Host Vehicle}
\newacronym{imsi}{IMSI}{International Mobile Subscriber Identity}
\newacronym{lcid}{LCID}{Logical Channel Identifier}
\newacronym{nnet}{NN}{Neural Network}
\newacronym{movav}{MovAv}{Moving Average}
\newacronym{idw}{IDW}{Inverse Distance Weighting}
\newacronym{lin}{Lin}{Linear Interpolation}
\newacronym{relu}{ReLU}{Rectified Linear Unit}
\newacronym{adam}{Adam}{Adaptive moment estimator}
\newacronym{drl}{DRL}{Deep Reinforcement Learning}
\newacronym{fcd}{FCD}{Floating Car Data}
\newacronym{csv}{CSV}{Comma Separated Values}
\newacronym{ca}{CA}{Carrier Aggregation}
\newacronym{nwdaf}{NWDAF}{Network Data Analytics Function}
\newacronym{etsi}{ETSI}{European Telecommunications Standards Institute}

\usepackage[capitalize]{cleveref}
\crefname{section}{Section}{Sections}
\crefname{figure}{Figure}{Figures}

\usepackage{tikz}
\usepackage{pgfplots}
\usepackage{tikzscale}
\usepackage{tikz-qtree}
\usepackage{multirow}

\usepackage{color, colortbl}
\definecolor{Gray}{gray}{0.9}

\DeclareFontFamily{\encodingdefault}{\ttdefault}{\hyphenchar\font=`\-}

\AtBeginDocument{%
  \providecommand\BibTeX{{%
    \normalfont B\kern-0.5em{\scshape i\kern-0.25em b}\kern-0.8em\TeX}}}

\begin{document}
\pagestyle{empty} 

\title{Artificial Intelligence in Vehicular Wireless Networks: \\ A Case Study Using ns-3}



\author{Matteo Drago}
\email{dragomat@dei.unipd.it}
\affiliation{%
  \institution{Dep. of Information Engineering\\University of Padova}
  \city{Padova}
  \state{Italy}
}

\author{Tommaso Zugno}
\email{tommaso.zugno@huawei.com}
\affiliation{%
  \institution{Huawei Technologies\\Munich Research Center}
  \city{Munich}
  \state{Germany}
}

\author{Federico Mason}
\email{masonfed@dei.unipd.it}
\affiliation{%
  \institution{Dep. of Information Engineering\\University of Padova}
  \city{Padova}
  \state{Italy}
}

\author{Marco Giordani}
\email{giordani@dei.unipd.it}
\affiliation{%
  \institution{Dep. of Information Engineering\\University of Padova}
  \city{Padova}
  \state{Italy}
}

\author{Mate Boban}
\email{mate.boban@huawei.com}
\affiliation{%
  \institution{Huawei Technologies\\Munich Research Center}
  \city{Munich}
  \state{Germany}
}

\author{Michele Zorzi}
\email{zorzi@dei.unipd.it}
\affiliation{%
  \institution{Dep. of Information Engineering\\University of Padova}
  \city{Padova}
  \state{Italy}
}


\begin{abstract}

\Gls{ai} techniques have emerged as a powerful approach to make wireless networks more efficient and adaptable.
In this paper we present an ns-3 simulation framework, able to implement \gls{ai} algorithms for the optimization of wireless networks.
Our pipeline consists of: (i) a new geometry-based mobility-dependent channel model for V2X; (ii) all the layers of a 5G-NR-compliant protocol stack, based on the \texttt{ns3-mmwave} module; (iii) a new application to simulate V2X data transmission,
and (iv) a new intelligent entity for the control of the network via AI.
Thanks to its flexible and modular design, researchers can use this tool to implement, train, and evaluate their own algorithms in a realistic and controlled environment. 
We test the behavior of our framework in a \gls{pqos} scenario, where AI functionalities are implemented using \gls{rl}, and demonstrate that it promotes better network optimization compared to baseline solutions that do not implement AI.

\begin{picture}(0,0)(0,-420)
\put(0,0){
\put(0,0){This paper has been submitted to WNS3 2022. Copyright may be transferred without notice.}}
\end{picture}

\end{abstract}



\maketitle

\glsresetall

\section{Introduction} 
\label{sec:introduction}
\Gls{ai} will be a key component of future \gls{6g} wireless networks~\cite{giordani2020toward}, as a means to achieve autonomous network optimization~\cite{letaief2019roadmap}.
In particular, the co-design of communication systems and applications with \gls{ai} in mind will allow 6G networks to learn, adapt, and support diverse services and requirements, without human~intervention.

Among other areas, \gls{ai} has been recognized as a promising technology in \gls{v2x} networks, to enable applications like traffic flow and congestion control, localization, platoon management, and autonomous driving~\cite{tong2019artificial}.
For these systems to be truly autonomous, intelligent vehicles need to acquire, process, and eventually disseminate massive amounts of data generated by on-board sensors~\cite{zhang2018vehicular}. 
Notably, \gls{ai}, in combination with \gls{ml}, can be designed to extract features from input data~\cite{giordani2019investigating}, and support complex V2X tasks like object detection and recognition~\cite{rossi2021role}, data compression~\cite{nardo2022point}, as well as tracking and trajectory prediction~\cite{baek2020vehicle}. 

As far as AI/ML is concerned, the availability of data for training and optimization is essential. 
In this regard, experiments with real testbeds are impractical due to limitations in the scalability and flexibility of these platforms, as well as the high cost of hardware components. 
On the other hand, AI-based research should follow an iterative approach where modeling and validation will be cyclically performed.
Therefore, computer simulations have emerged as important tools for testing the performance of AI solutions in different conditions and scenarios. 
Python-based simulators are among the most popular softwares for AI, thanks to desirable features like versatility and readability~\cite{nagpal2019python}. 
Most importantly, many open-source libraries providing base-level ready-to-use coding solutions to develop AI functionalities, like TensorFlow, PyTorch or Keras, are implemented in Python.
When it comes to the simulation of wireless networks, however, Python simulators tend to rely on simplified assumptions on the system architecture, and have not proven particularly successful in modeling the protocol stack of complex networks. 
Instead, discrete-event network simulators, like ns-3~\cite{henderson2008network}, are valid alternatives to analyze the performance of wireless networks in more realistic scenarios.
How to integrate the two simulators and build an open software able to support end-to-end full-stack simulations, as well as network optimization via AI, is still an open challenge. 
An initial effort in this direction was made in~\cite{piotr2019ns3meetsgym}, even though the authors focused on OpenAI Gym, a toolkit specific for \gls{rl} research. 

In this paper, we fill this gap and propose a simulation pipeline to design and test AI in wireless networks. 
In particular, we focus on a V2X scenario, in which teleoperated vehicles use a cellular connection to exchange sensor data with a remote driver.
Specifically, we extended the ns-3 \texttt{mmwave} module~\cite{mezzavilla2018end}, one of the most 5G-oriented frameworks to simulate wireless networks, as follows:
\begin{itemize}
  \item We implemented a geometry-based channel model for V2X, based on GEMV$^2$ and \gls{sumo} traces, that is consistent with the actual deployment of buildings and vehicles in the scenario.
  \item We designed a new ns-3 application to simulate the traffic flow in V2X use cases. Specifically, the ns-3 application involves the exchange of sensor data, which may be preemptively compressed and/or segmented to reduce the file size before transmission, modeled based on the Kitti multi-modal dataset~\cite{geiger2012are}. The application is characterized by (i) the size of the input data, and (ii) the time periodicity at which the information is generated and exchanged, and (iii) the level of compression/segmentation of the data.
  \item We introduced a new entity called ``RAN-AI'' (the core contribution of this paper) that, connected to the \gls{ran} and with Python-based AI algorithms, optimizes network operations. 
\end{itemize}

As a case study, we validate our AI pipeline for a \gls{pqos} application~\cite{boban2021predictive}, defined as a mechanism to predict \gls{qos} changes and provide autonomous vehicles with advance notifications to react accordingly. 
In this scenario, the RAN-AI collects full-stack network metrics from different components of the RAN, and implements an \gls{rl} framework (first introduced in~\cite{mason2022rlpqos}) that is able to identify the optimal network configuration.
We demonstrate that our RAN-AI is able to improve the \gls{qos} of V2X applications, compared to other baseline solutions that do not implement \gls{ai} techniques.

The rest of the paper is structured as follows.
In Sec.~\ref{sec:ns_3_implementation} we discuss how we integrated AI operations in ns-3.
In Sec.~\ref{sec:results} we validate our ns-3 implementation for PQoS.
In Sec.~\ref{sec:conclusions} we conclude the paper with suggestions for future research.

\begin{figure*}[t!]
 \centering
   \includegraphics[width=0.8\textwidth]{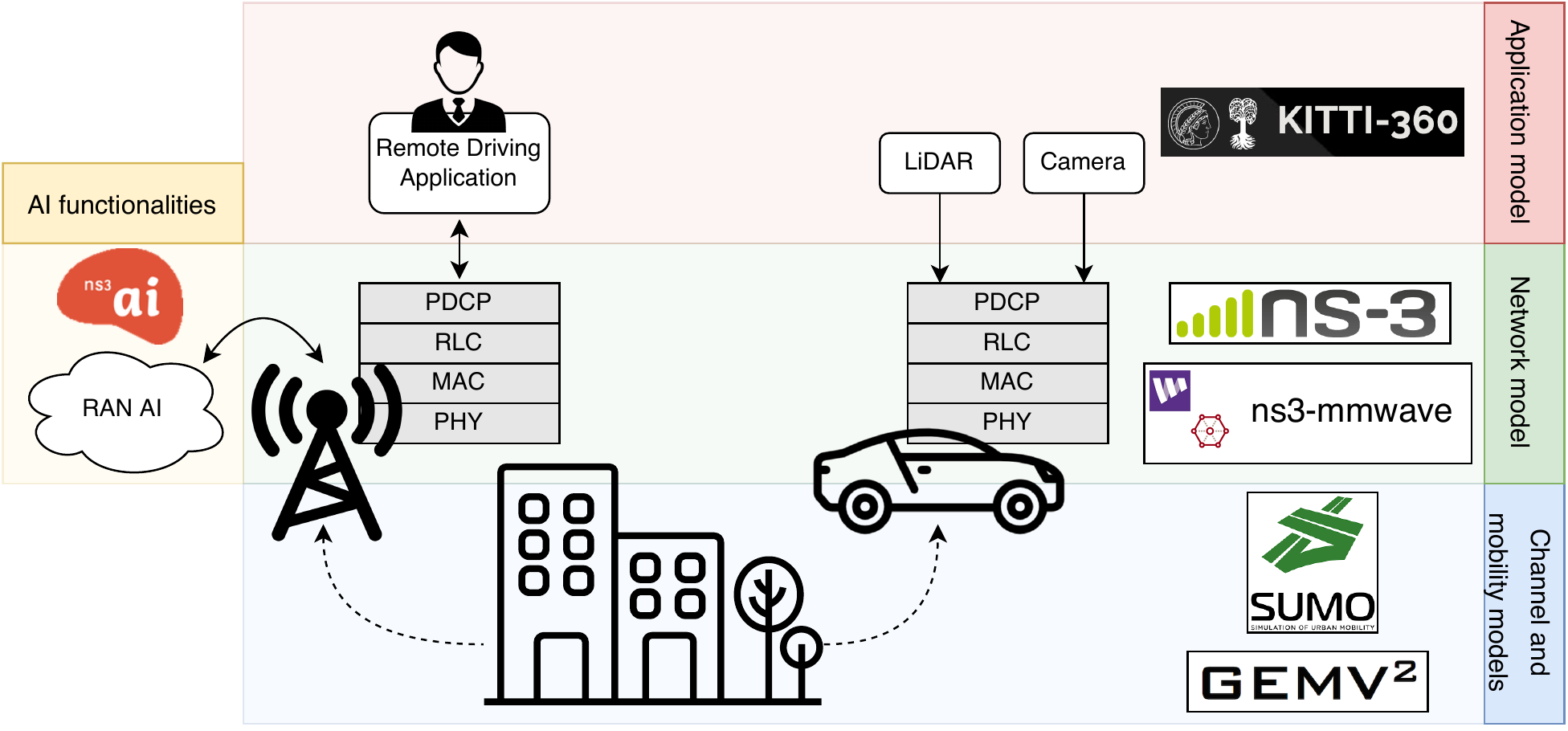}
   \caption{Overview of the proposed ns-3 simulation framework/pipeline to incorporate AI functionalities in wireless networks.}
   \label{fig:overview}
\end{figure*}

\section{Design of intelligent wireless networks using ns-3} 
\label{sec:ns_3_implementation}


Figure~\ref{fig:overview} provides an overview of the ns-3 simulation framework that we developed to integrate AI functionalities in vehicular networks. 
We distinguish four main components, namely (i) the channel and mobility models (Sec.~\ref{sub:channel}), 
(ii) the network model (Sec.~\ref{sub:network}), which simulates the communication network, (iii) the application model (Sec.~\ref{appModel}), which mimics a real vehicular application, and (iv) the intelligent network controller (Sec.~\ref{sub:ran-ai}), which provides \gls{ai} functionalities to optimize the network configuration. 
The source code is publicly available at~\cite{ns3-ran-ai}.


\subsection{Channel Model}
\label{sub:channel}
A realistic characterization of the wireless channel is of paramount importance to obtain accurate simulation results~\cite{lecci2021accuracy}. The channel model should incorporate the impact of the radio environment on the propagation of wireless signals (e.g., the presence of buildings and other blockers) and user mobility.  

The approach adopted in our framework relies on \gls{osm} to obtain a detailed representation of the area of interest. 
The generation of vehicles' mobility traces is handled by \gls{sumo}~\cite{SUMO2012}, a popular open-source tool for the simulation of vehicular traffic.
In particular, the \gls{osm} representation of the scenario is converted into a \gls{sumo} network file using the \texttt{netconvert} utility, and the generation of random routes is carried out using the script \texttt{randomTrips.py}.
Finally, the wireless channel is computed using GEMV$^2$, a geometry-based propagation model for \gls{v2x} scenarios~\cite{boban2014geometry}, which is open source and publicly available~\cite{gemv2}.
GEMV$^2$ calculates both geometry-based large- and small-scale fading components of the channel from a map of the environment and the trajectories of the vehicles, taking into account the effect of mobility, buildings outlines and foliage, and outputs the propagation loss for all the possible device pairs at each time step. 

  
To feed the channel traces into ns-3, we created a parser able to read the GEMV$^2$ output files, and eventually compute the received power between a pair of devices based on the current value of the transmit power. 
The parser is implemented by the class \texttt{Gemv\-Pro\-pa\-ga\-tion\-Loss\-Mo\-del}, which exploits the modular design of the ns-3 \texttt{pro\-pa\-ga\-tion} module by extending the \texttt{Pro\-pa\-ga\-tion\-Loss\-Mo\-del} interface.  
The new class provides the method \texttt{Do\-Calc\-Rx\-Power()}, which reads the traces and retrieves the current simulation time to determine the received power for the desired link(s). 
For an efficient processing of the \gls{csv} files, we used the \texttt{Csv\-Reader} class available in the ns-3 \texttt{core} module.

\subsection{Network Model}
\label{sub:network}
For an accurate characterization of the network components, we chose to extend the \texttt{ns3-mmwave} module, an open source an publicly available ns-3 module for the simulation of 5G networks~\cite{mezzavilla2018end}.  

The \texttt{ns3-mmwave} module implements models for all the layers of the 5G~NR protocol stack, for both \glspl{gnb} and \glspl{ue}.
The custom \gls{phy} and \gls{mac} layers support different NR-compliant frame structures and numerologies, multiple beamforming algorithms and scheduling policies. The \gls{rlc} and \gls{pdcp} layers, as well as the core network models, are based on the ns-3 \texttt{lena} module for \gls{lte} networks~\cite{piro2011lte}. 
It also supports dual connectivity with \gls{lte} base stations, which enables the simulation of non-standalone 5G deployments, and \gls{ca} at the \gls{mac} layer. 
Based on these features, \texttt{ns3-mmwave} supports end-to-end full-stack network simulations with a high level of detail, and has been taken as a reference benchmark simulator for 5G scenarios.

Although this module is meant for the simulation of 5G systems operating at \gls{mmwave} frequencies, we introduced some modifications to make it work at lower frequencies too. 
In particular, we modified the class \texttt{Mm\-Wa\-ve\-Hel\-per} to accept the GEMV$^2$-based propagation model presented in Sec.~\ref{sub:channel}, and create \glspl{gnb} and \glspl{ue} without beamforming capabilities (via the \texttt{In\-stall\-Sub\-6\-Gnb\-De\-vi\-ce()} and \texttt{In\-stall\-Sub\-6\-Ue\-De\-vi\-ce()} methods, respectively). 

\subsection{Application Model}
\label{appModel}

We focused on a teleoperated driving scenario where a \gls{hv}
is controlled by a remote driver through an ad hoc driving application installed on a remote or edge server.
In order to be teleoperated, the \gls{hv} must be able to disseminate perception data generated from on-board sensors like \glspl{lidar} to the remote driver, that will then detect/recognize sensitive entities in the environment (e.g., cars, pedestrians, cyclists, etc.). 
However, the transmission of sensor data requires a considerable amount of radio resources and could potentially congest the network, preventing a smooth driving experience. To tackle this issue, data compression and segmentation are often applied in order to reduce the size of raw data prior to transmission.


In ns-3, we designed a new application module to simulate this data exchange and generate sensor data as a function of:
\begin{enumerate}
    \item The size of the input sensor data (in bytes);
    \item The time periodicity at which the information is generated and exchanged (typically fixed to 100 ms for \gls{lidar} sensors, according to the device's data sheets~\cite{velodyne});
    \item The level of compression/segmentation for the sensor data.
\end{enumerate}

In this work we consider the sensor data from the Kitti multi-modal dataset, collected using a Volkswagen Passat equipped with a Velodyne \gls{lidar}
~\cite{geiger2012are}. 
In particular, we rely on the data compression pipeline proposed in~\cite{varischio2021hybrid}.
First, we infer semantic segmentation of point clouds with RangeNet++~\cite{rnet}. We consider 3 segmentation~levels:
\begin{itemize}
    \item \emph{Raw (R)}: The raw LiDAR acquisition is considered.
    \item \emph{Segmentation Conservative (SC)}: Data points associated to road elements are removed, thus reducing the file size.
    \item \emph{Segmentation Aggressive (SA)}: Data points associated to buildings, vegetation, traffic signs, and the background are also removed, thus keeping only the most critical items in the scene (typically pedestrians and vehicles).
\end{itemize}
Second, we compress the resulting frame using Draco~\cite{Draco}, a software whose flexibility allows the support of 15 quantization levels and 11 compression levels. 
Since there are 3 segmentation levels, 15 quantization levels and 11 compression ratios, overall there would be $495$ distinct alternatives; for simplicity, our application implements the 7 most representative ones, in terms of the trade-off between compression accuracy and speed, referred to as ``application modes'' in the rest of the paper. 

To implement the features described above, we started from the application presented in \cite{lecci21bursty}, able to generate large data frames and automatically fragment them into bursts of packets that are then re-aggregated at the receiver, whenever possible.
Specifically, we extended the \texttt{TraceFileBurstGenerator}, that allows the user to reproduce real world traffic traces, into the \texttt{Kit\-ti\-Tra\-ce\-Bur\-st\-Gen\-era\-tor}.
Using the \texttt{CsvReader} utility, already available in ns-3, we created a method to import sensor traces from the Kitti dataset (after applying compression and segmentation to the data where applicable), and save each frame information into a data structure.

As in a specific time instant of the simulation the application can operate in one specific application mode, we provided \texttt{Kit\-ti\-Tra\-ce\-Bur\-st\-Gen\-era\-tor} with ad hoc methods to change the mode in real time.
In addition, considering that a single traffic trace could incorporate and/or last for multiple scenes, the user can choose what sequence the application is replicating and decide, in case the reader reaches the end of the scene of interest, whether to loop again from the beginning of the same scene or stop the application.

Considering that a \gls{lidar} collects 3D points periodically, we also emulate this behavior by the design of the \texttt{FramePeriod} attribute, that indicates the time interval between a frame and the following one, and is used to subsequently schedule the sending of each burst.
The \texttt{BurstyApplication} and \texttt{BurstSink} classes will then take care of burst fragmentation, transmission and reception of packets, and re-aggregation of the burst. 

For easy collection of statistics at the application, the module was integrated with an additional \texttt{\hyphenchar\font=`\- BurstyAppStatsCalculator} utility class, which creates an output file to report how many bytes and bursts were received in a specific window of time, and the delay the received bursts have accumulated, on average, for each node.

\subsection{Intelligent Network Controller}
\label{sub:ran-ai}

As highlighted in Sec.~\ref{sec:introduction}, the main goal of this work is to develop a framework to integrate AI functionalities in ns-3. 
In this section, we present a new entity called RAN-AI, installed at the \gls{gnb}, which interfaces with different components of the \gls{ran}, as well as the core network, and incorporates \gls{ai} capabilities with the purpose of optimizing V2X network operations.
In particular, the RAN-AI collects network metrics and takes actions (also referred to as ``countermeasures'') to control the connected vehicles accordingly. 

In our framework, the RAN-AI is responsible for:
\begin{enumerate}
  \item Collecting metrics from the gNB (i.e., cell-related information) and the end users.
  \item Running AI algorithms using as inputs the collected metrics. We highlight that our implementation is AI-agnostic, in the sense that it supports different AI models able to solve heterogeneous problems. 
  \item Determining the actions to take in order to maximize the network performance.
  \item Communicating the actions to the relevant entities so that they can tune their behavior accordingly. In particular, in this work we allow the RAN-AI to control the end users, even though our framework does not prevent other countermeasures to be considered.
\end{enumerate}

With respect to the code structure, RAN-AI functionalities are implemented by the \texttt{RanAI} class, that is in charge of integrating the features of \texttt{ns3-ai}~\cite{hao2020ns3ai}, an ad hoc module to provide efficient and high-speed data exchange between Python-based AI algorithms and ns-3. 
Specifically, this module uses a shared memory implementation for interprocess communications, and provides a high-level interface in both Python and C++ for different algorithms, which is the reason why we integrated it in our framework.

To enable this new entity, an instance of the \texttt{RanAI} has to be installed on each gNB in the simulation environment.
First, we included in \texttt{MmWaveEnbNetDevice} the attribute \texttt{m\_ranAI} representing the RAN-AI instance, which is initialized by calling the class method \texttt{InstallRanAI()}. 
Specifically, \texttt{InstallRanAI()} (i) initializes instances that gather full-stack network statistics through the classes \texttt{Mm\-Wave\-Bearer\-Stats\-Cal\-cu\-la\-tor} and \texttt{Bur\-sty\-App\-St\-ats\-Cal\-cu\-la\-tor}, and (ii) schedules \texttt{Send\-Sta\-tus\-Up\-da\-te()}, a routine that is executed every \texttt{m\_statusUpdate} seconds.
In particular, \texttt{SendStatusUpdate()} allows:
\begin{enumerate}
  \item The gNB to collect measurements at the \gls{phy}, \gls{mac}, \gls{rlc}, \gls{pdcp}, and application layers, pertaining to the end users.
  \item The gNB to organize this information to be compatible with \texttt{Re\-port\-Mea\-sures()} input requirements and, through this method, provide it to the RAN-AI.
  \item The RAN-AI to process the dataset via the \gls{ai} framework. 
\end{enumerate}

When it comes to delivering the agent's decision (i.e., the optimal action) to an end user, the RAN-AI checks whether that user is already configured to operate as specified by the action.
If not, the action must be communicated to the end user, and in our ns-3 framework we devise two possibilities:
\begin{itemize}
  \item \emph{Ideal notification}: We directly trigger a callback function to apply the agent's decision.
  No packet is transmitted, thus we do not model the impact of the transmission delays and/or communication errors/failure when changing the action.

  \item \emph{Real notification}: The notification involves the transmission of a real packet towards the intended end user. Specifically, each notification packet contains (i) the agent's decision, (ii) the \gls{imsi} of the user, and (iii) the \gls{rnti} of the user in the cell. 
  As such, both transmission delays and communication errors are involved in the process (i.e., the packet could be lost, leading the end user to operate sub-optimally).\footnote{We recall that our framework is based on the \texttt{ns3-mmwave} module, that we extended to support communication at both sub-6 GHz and \gls{mmwave} frequencies.}
\end{itemize}

With respect to the AI framework running on top of the RAN-AI, we deployed multiple \glspl{mdp}, each of which represents the behavior of a distinct end user. 
Specifically, we developed a novel {Python} module, named \texttt{CentralizedAgent}, which allows an agent to interact with multiple learning environments, and to test multiple learning configurations, thereby evaluating the performance of different AI algorithms applied to the same scenario. 
The \texttt{CentralizedAgent} module implements two main methods, namely \texttt{get\_action()} and \texttt{update()}.
\begin{itemize}
  \item \texttt{get\_action()} takes as an input the states of the $N_u$ end users in the target scenario, and computes $N_u$ actions.
  Notably, the state of an end user is a vector that includes all the input parameters that the RAN-AI entity can collect for the given vehicle. 
  Depending on how state and action spaces are explored, the method will return different~actions.
  \item \texttt{update()} takes as an input a list of $N_u$ \emph{learning transitions}, i.e., sequences [$s_t$, $a_t$, $s_{t+1}$, $r_t$] associated with each of the $N_u$ end users in the target scenario during slot $t$. 
Specifically, a learning transition consists of (i) the state $s_t$ of the end user at the beginning of time slot $t$, (ii) the action $a_t$ performed by the end user during slot $t$, (iii) the state $s_{t+1}$ at the beginning of the next slot $t+1$ and, (iv) the reward $r_t$ that the end users receives at the end of slot $t$.
In general, the quality of the agent's decisions increases as more data are gathered by the \texttt{update()} method.
\end{itemize}
The reward function is specific to the target application. As a case study, in this paper we will validate our AI framework for a \gls{pqos} application, as described in Sec.~\ref{sub:learning_setup}. 
Nevertheless, our RAN-AI implementation is transparent to the ns-3 scenario, the only dependence being how the input data is structured.

It should be mentioned that training an AI algorithm may require a huge computational effort.
In our case, this challenge is exacerbated by the fact that the algorithm's input is directly taken from the ns-3 simulation running under the hood, to provide information on how the system evolves and reacts to the decisions of the agent. 
At every step, in fact, the agent needs to wait for ns-3 to compute and collect a set of metrics, that are then used for training. 
In this sense, the proposed framework supports \emph{transfer learning}, where initial learning can be performed in a simplified (and faster) environment, with further training in a more realistic environment (in our case, in ns-3).

\section{A Case Study} 
\label{sec:results}
In this section, we validate our AI simulation pipeline for a \gls{pqos} use case, and demonstrate that it represents a valid tool to evaluate and optimize wireless networks. 
In Sec.~\ref{sub:learning_setup} we present our AI learning setup, in Sec.~\ref{sub:simulation_parameters} we describe our simulation parameters, and in Sec.~\ref{sub:simulation_results} we show initial numerical results.

\subsection{AI Algorithm} 
\label{sub:learning_setup}
\paragraph{Overview}
As a case study, we focus on \gls{pqos}, a paradigm to provide autonomous systems with advanced notifications in case of upcoming \gls{qos} changes~\cite{boban2021predictive}.
To address this problem, the RAN-AI introduced in Sec.~\ref{sub:ran-ai} collects network statistics at the RAN level and, based on them, defines and applies network countermeasures in case QoS requirements are not satisfied. 
Specifically, we designed the RAN-AI to implement an AI agent, based on \gls{rl}, able to identify the optimal application mode (see Sec.~\ref{appModel}) for the end users when transmitting sensor data. 
The rationale behind this choice is that end users will be encouraged to select more aggressive application modes (e.g., those that apply compression/segmentation) to reduce the size of the packets to send and promote faster transmissions.

\paragraph{Learning agent} 
Based on our previous work~\cite{mason2022rlpqos}, our AI/RL agent is trained according to the \gls{dql} algorithm described in \cite{van2016deep}, which is an extended version of the classical \emph{Q-learning}~\cite{watkins1992q}.
Hence, whenever the \texttt{update()} function is called, \texttt{CentralizedAgent}
follows the \gls{dql} procedure to perform a new training step.
Our framework approximates the agent's policy by means of a \gls{nnet}, which makes it possible to handle continuous state spaces and overcome the \emph{curse of dimensionality} phenomenon~\cite{bellman1966dynamic}.
We consider a Feed-Forward \gls{nnet}, with $S$ inputs and $A$ output neurons, and implement the \gls{relu} activation function across the different layers~\cite{agarap2018deep}. 

We highlight that the input size of the \gls{nnet} coincides with the dimension of the system's state, i.e., the number of input parameters of the RAN-AI entity.
Instead, the output size of the \gls{nnet} corresponds to the number of possible actions for the agent, i.e., in our case the different application modes.

\paragraph{Reward function} 
In our target scenario, the performance of the system depends on two different aspects: 
\begin{itemize}
	\item The \acrfull{qos}: The vehicles should satisfy \gls{qos} requirements, especially in terms of maximum end-to-end delay $\delta_M$ and minimum \gls{prr} PRR$_m$.
	
	\item The \acrfull{qoe}: The transmitted data should be accurate enough to perform driving operations. For our case study, the \gls{qoe} depends on the symmetric point-to-point Chamfer Distance $\mathrm{CD}_{\rm sym}$, which is inversely proportional to the quality of the received data~\cite{varischio2021hybrid}.
\end{itemize}

To incorporate both these factors, the agent reward is designed to balance between \gls{qos} and \gls{qoe} via a tuning parameter $\alpha\in[0,1]$.

Let $\hat{\text{PRR}}_t$ and $\hat{\delta}_t$ be the \gls{prr} and average delay of the vehicle at time $t$, respectively.
If the \gls{qos} requirements are not met, i.e., $\hat{\delta}_t \geq \delta_M$ and $\hat{\text{PRR}}_t < \text{PRR}_m$, the agent reward $R(t)$ is $0$, otherwise it is given by
\begin{equation}
\label{eq:reward}
R(t) = (1-\alpha) \frac{\delta_M - \hat{\delta}_t}{\delta_M} + \alpha \frac{\mathrm{CD}_{\text{sym},m} - \hat{\mathrm{CD}}_{\text{sym},t}}{\mathrm{CD}_{\text{sym},m}},
\end{equation}
where $\hat{\mathrm{CD}}_{\text{sym},t}$ and $\mathrm{CD}_{\text{sym},m}$ are the Chamfer Distance at time $t$ and the maximum Chamfer Distance that can be tolerated,~respectively.

\begin{table}[t!]
\centering
\scriptsize
\renewcommand{\arraystretch}{1.3}
\caption{Simulation parameters.}
\label{tab:params}
\begin{tabular}{c|c|c}
  \toprule
  Parameter & Description & Value \\
  \hline
  
  $f_c$ & Carrier frequency & 3.5 GHz       \\ \hline
  $B$ & Total bandwidth                           & 50 MHz        \\ \hline
  $P_{TX}$ & Transmission power & 23 dBm        \\ \hline
  $T$ & RAN-AI update periodicity        & 100 ms        \\ \hline
  $\tau_s$ & Simulation time             & 80 s          \\ \hline
  $N_u$ & Number of vehicles & $\{1,\,5\}$ \\ \hline
  $\lambda$ & Discount factor  & 0.95          \\ \hline
  $\zeta$ & Learning rate      & $10^{-4}$     \\ \hline
  $\epsilon$ & Weight decay    & $10^{-3}$     \\ \hline
  $\alpha$ & QoS/QoE weight       & 1 \\ \hline
  $\delta_M$ & Max. tolerated delay    & $50$ ms     \\ \hline
  PRR$_m$ & Min. tolerated PRR    & 1   \\ \hline
  $\mathrm{CD}_{\text{sym}, m}$ & Max. tolerated Chamfer Distance  & 45     \\ \hline
  \multicolumn{2}{c|}{Layer size (inputs $\times$ outputs)} & $8\times 12 \rightarrow 12\times 6 \rightarrow 6\times 3$ \\ 
  \bottomrule
  \end{tabular}
\end{table}

\subsection{Simulation Parameters} 
\label{sub:simulation_parameters}

The simulation parameters are shown in Table~\ref{tab:params}.

\begin{figure}[t!]
 \centering
   \includegraphics[width=0.95\columnwidth]{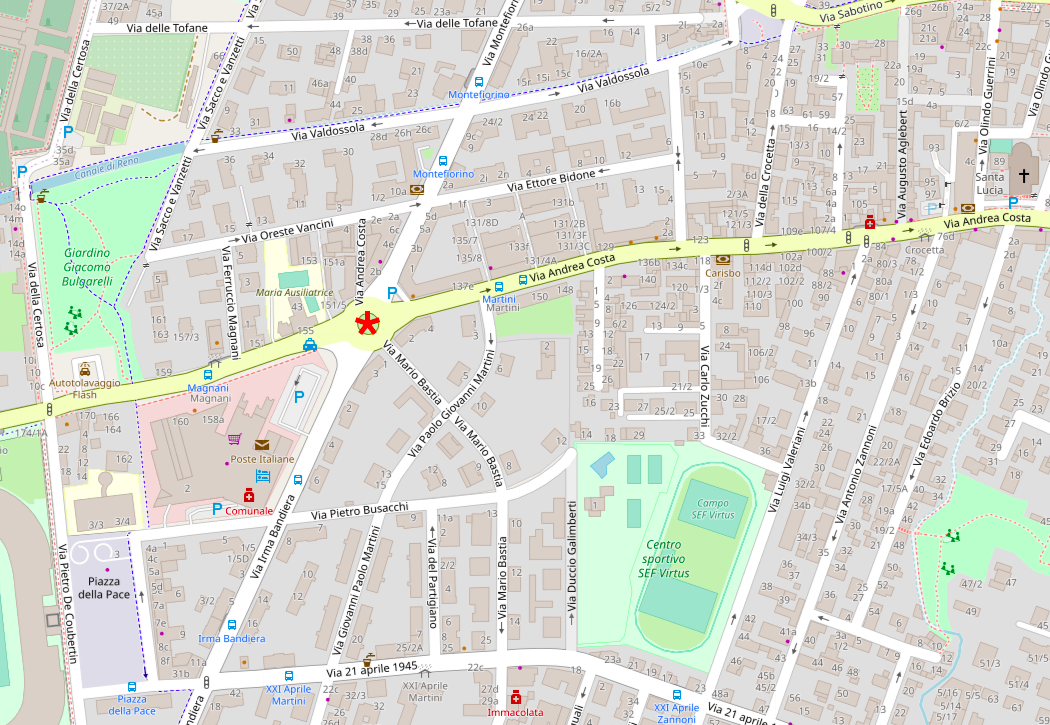}
   \caption{Our simulation scenario, corresponding to a portion of the city of Bologna (Italy). The red star represents the position of the gNB.}
   \label{fig:bologna}
\end{figure}

\paragraph{Scenario}
We consider a scenario with $N_u$ teleoperated vehicles traveling on a real road topology, which corresponds to the portion of the city of Bologna (Italy) depicted in Figure~\ref{fig:bologna}. 
There are two main streets connected by a circular intersection, and several urban and sub-urban streets. 
This area includes both commercial and residential buildings with different heights and sizes. 
We consider a single \gls{gnb}, operating at 3.5~GHz with a bandwidth of 50~MHz, at the center of the circular intersection at a height of 6.5~m (represented with a red star in the figure).

\paragraph{V2X application}
Each vehicle runs an instance of the \texttt{KittiApplication} presented in Sec.~\ref{appModel} for streaming LiDAR data to a remote driver, and receives downlink commands for teleoperated driving operations, which is modeled as a UDP source with constant rate 0.32 Mbps.

\paragraph{AI/RL algorithm}
The RAN-AI collects network metrics every 100~ms, and implements the RL algorithm described in Sec.~\ref{sub:learning_setup} to optimize network operations. 
In our implementation, the agent's action space $\mathcal{A}$ is limited to four most representative actions, corresponding to the set of application modes $\mathcal{A} \in$ \{C-R, C-SC, C-SA\}, where C-R means that data are compressed (with compression level $14$) but not segmented, C-SC that data is compressed and conservative segmentation is applied, and C-SA that data is compressed and aggressive segmentation is~applied. 
In terms of the reward function in Eq.~\eqref{eq:reward}, we set the tuning parameter to $\alpha=1$ (i.e., the agent tries to maximize the QoE, as long as QoS demands are satisfied), while communication requirements for teleoperated driving are based on 5GAA specifications~\cite{5gaa2020cv2x}, so we have $\delta_M=50$ ms and PRR$_m=1$. Finally, $\mathrm{CD}_{\text{sym}, m}$ is set to 45, while $\hat{\mathrm{CD}}_{\text{sym},t}$ depends on the application mode and increases as considering more aggressive compression/segmentation.
Simulation results as a function of $\alpha$ can be found in~\cite{mason2022rlpqos}, where we proved that decreasing $\alpha$ has the benefit to further improve the QoS, even beyond the QoS requirements under consideration, at the expense of some QoE degradation.


\subsection{Results} 
\label{sub:simulation_results}

To validate our framework, we compared the following~policies:
\begin{itemize}
  \item \emph{DQL} (proposed), where at each step the agent implements our proposed RAN-AI framework.
  \item \emph{Constant} (benchmark), where at the beginning of the simulation the end user maintains one application mode $\in\mathcal{A}$ for the whole simulation. 
\end{itemize}
The two policies have been tested separately, and will be compared in terms of the user's (i) \gls{qos}, expressed in terms of delay and PRR at the application layer, and (ii) \gls{qoe}, which depends on the Chamfer Distance.
We investigate the impact of the RAN-AI implementation for delivering the agent's decisions to the end users (real or ideal) and the number of vehicles $N_u$.


\begin{table}[t!]
\centering
\scriptsize
\renewcommand{\arraystretch}{1.3}
\caption{Average QoE $\in[0,1]$ performance of different PQoS policies, as a function of the RAN-AI implementation for notifications (real or ideal) and~$N_u$.}
\label{tab:qoe}
\begin{tabular}{>{\centering\arraybackslash} p{0.15\columnwidth}|>{\centering\arraybackslash} p{0.15\columnwidth}|>{\centering\arraybackslash} p{0.15\columnwidth}|>{\centering\arraybackslash} p{0.15\columnwidth}|>{\centering\arraybackslash} p{0.15\columnwidth}}
  \toprule
  \multirow{2}{*}{PQoS policy} & \multicolumn{2}{c|}{\shortstack{RAN-AI notification \\ ($N_u=1$)}} & \multicolumn{2}{c}{\shortstack{Number of vehicles~$N_u$ \\ (Real Notification)}} \\
  \cline{2-5}
  & Real & Ideal & $1$ & $5$ \\  \hline
  C-R & 1 & 1 & 1 & 1  \\ \hline
  C-SC & 0.88 & 0.88 & 0.88 & 0.88  \\ \hline
  C-SA & 0.22 & 0.22 & 0.22 & 0.22 \\ \hline
  \rowcolor{Gray} DQL & 0.98 & 0.94 & 0.98 & 0.78 \\
  \bottomrule
  \end{tabular}
\end{table}

\begin{figure}[t!]
  \centering
  \begin{subfigure}{.99\linewidth}
    \centering
    \includegraphics[width=\linewidth]{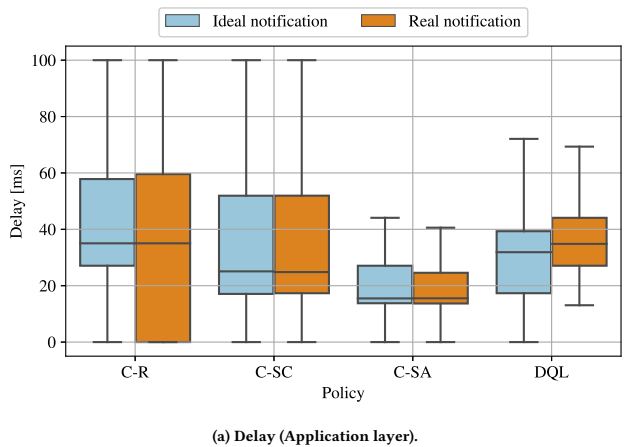}
    \caption{Delay (Application layer).}
    \label{fig:multi_update_delay}
  \end{subfigure}\vspace{0.5cm}
  \begin{subfigure}{.99\linewidth}
    \centering
    \includegraphics[width=\linewidth]{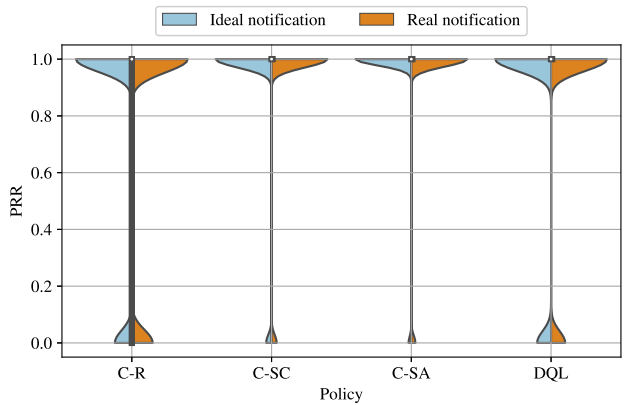}
    \caption{PRR (Application layer).}
    \label{fig:multi_update_prr}
  \end{subfigure}
  \caption{Performance of different PQoS policies with $N_u=1$, considering the impact of the RAN-AI overhead for notifications.}
  \label{fig:multi_update_comm}
\end{figure}

In Table~\ref{tab:qoe} we report the \gls{qoe} for different PQoS policies. 
We observe that all the baseline solutions show constant values of the QoE, regardless of the RAN-AI implementation and $N_u$, as the compression level remains always fixed, while \gls{dql} shows different \gls{qoe} performance depending on the adopted policy. The best \gls{qoe} is achieved when transmitting raw sensor data (C-R), given that data segmentation privileges efficiency over accuracy and eventually distorts the LiDAR data before transmission. 
Moreover, from Table~\ref{tab:qoe} we can see that introducing ideal notifications leads to QoE degradation, on average from 0.98 to 0.94.
Although this behavior may sound counterintuitive, we found that, when using ideal notifications, \gls{dql} was encouraged to adopt a more aggressive behavior, i.e., the agent was encouraged to change the application mode more frequently throughout the learning process. This has the benefit to improve the QoS (as we will show in Fig.~\ref{fig:multi_update_comm}), at the cost of a lower QoE.
This is an indication that QoS and QoE should be studied together to really understand the performance of the system, as we will discuss in the following results.


\begin{figure}[t!]
  \centering
  \begin{subfigure}{.99\linewidth}
    \centering
    \includegraphics[width=\linewidth]{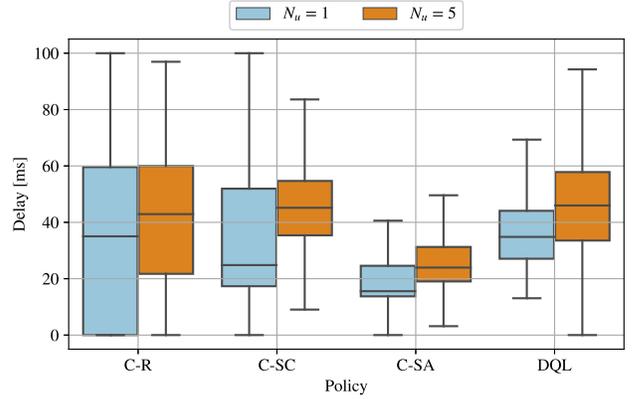}
    \caption{Delay (Application layer).}
    \label{fig:multi_user_delay}
  \end{subfigure}\vspace{0.5cm}
  \begin{subfigure}{.99\linewidth}
    \centering
    \includegraphics[width=\linewidth]{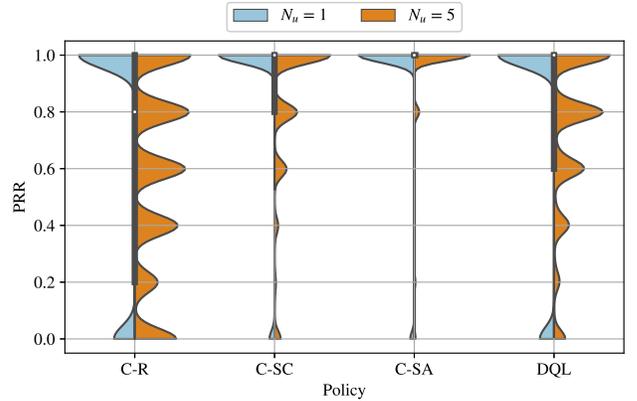}
    \caption{PRR (Application layer).}
    \label{fig:multi_user_prr}
  \end{subfigure}
  \caption{Performance of different PQoS policies, considering \emph{real notifications} at the RAN-AI and the impact of the number of users.}
  \label{fig:multi_user_comm}
\end{figure}

In Figures~\ref{fig:multi_user_delay} and \ref{fig:multi_user_prr} we illustrate the distributions of the delay and \gls{prr}, respectively, experienced at the application layer vs. $N_u$. 
Precisely, Figure~\ref{fig:multi_user_prr} represents a \emph{split violin plot}, which depicts the estimated probability density function of the \gls{prr} for different values of $N_u$, for each application mode.

First, as expected, the median and the percentiles of the delay (PRR) are always higher (lower) when $N_u = 5$, due to the fact that increasing the number of vehicles may congest the communication channel, thus degrading the overall system performance. This result validates the accuracy and realism of our ns-3 framework.

In terms of QoS, C-SA outperforms any other solution, since in this configuration the data are extremely compressed and segmented before transmission, thus reducing the size of the packets to send, at the expense of a very low QoE (0.22). 
In turn, C-R maximizes the QoE, but results in a QoS degradation (up to $2.3\times$ higher latency compared to C-SA). 
It appears clear that, in our DQL implementation, the RAN-AI tries to adapt the compression level to the conditions of the scenario via AI/RL, and achieves the best trade-off between QoE and QoS. 
In particular, with $N_u=1$, \gls{dql} is able to guarantee an average \gls{qoe} of 0.98, while ensuring an average end-to-end delay lower than 40~ms and a reliable data delivery. 
With $N_u=5$, \gls{dql} sacrifices the performance in terms of \gls{qoe} (0.78) in order to ensure an average delay lower than the maximum tolerated value for teleoperated applications (i.e., 50~ms). Furthermore, \gls{dql} reduces the variability of both the delay and the PRR compared to its competitors, a critical requirement in~V2X.


In Figure \ref{fig:multi_update_comm} we demonstrate that the impact of the additional overhead introduced when control notifications are transmitted from the RAN-AI to the end users  is not negligible. 
Specifically, with \emph{ideal notification} settings, both the delay and the \gls{prr} improve: the median delay and percentiles take lower values, while the probability of packet loss decreases, given that the transmission of real notification packets may incur additional delays and communication errors. 

\section{Conclusions} 
\label{sec:conclusions}
In recent years, V2X networks are incorporating \gls{ai} as a method to analyze large volumes of data and self-optimize. 
While simulators like ns-3 have been popular tools to analyze the performance of wireless networks, how to simulate \gls{ai} techniques and their impact on the communication stack is still an open question.
In this paper we address this challenge, and propose a novel framework in ns-3 able to simulate \gls{ai} algorithms.
To this aim, we implemented a new geometry-based channel model and application for V2X, and a new intelligent entity (called RAN-AI) for optimizing wireless networks.
We demonstrate the accuracy and technical soundness of our framework in a test scenario where the network performance is controlled via \gls{pqos}. We show from ns-3 simulations that V2X performance requirements in terms of QoS and QoE can be satisfied when the RAN-AI implements an RL algorithm for optimization.
We provide the source code of the simulator at~\cite{ns3-ran-ai}, in the hope that it will be useful to the broader community when implementing and evaluating AI techniques to improve wireless networks performance.

\bibliographystyle{ACM-Reference-Format}
\bibliography{bibl.bib}

\end{document}